\documentclass[11pt,reprint,english,aps,prc,showpacs,letter,longbibliography]{revtex4-1}

\usepackage[     bookmarks,
                 bookmarksopen = true,
                 bookmarksnumbered = true,
                 linktocpage,
                 colorlinks = true,
                 linkcolor = blue,
                 urlcolor  = blue,
                 citecolor = blue,
                 anchorcolor = green,
                 hyperindex = true,
                 hyperfigures]
        {hyperref}

\usepackage{graphicx}
\usepackage{dcolumn}
\usepackage{bm}
\usepackage{float}
\usepackage{amsmath}

\usepackage{color}
\usepackage{graphicx}
\definecolor{dgreen}{cmyk}{1.,0.,1.,0.2}        
\definecolor{orange}{cmyk}{0.,0.353,1.,0.}    

\newcommand{\trento}{T\raisebox{-.5ex}{R}ENTo}

\newcommand\sect[1]{\section{#1}}

\begin{document}

\title{A Novel Deep Learning Method for Detecting Nucleon-Nucleon Correlations}

\author{Yu-Jing Huang$^{1,2}$,  Zhu Meng$^{1,2}$, Long-Gang Pang$^{1*,2}$\footnote{email: lgpang@ccnu.edu.cn}, 
Xin-Nian Wang$^{1*,2}$\footnote{email: xnwang@lbl.gov}
}

\address{$^{1}$ Key Laboratory of Quark and Lepton Physics (MOE) and Institute of Particle Physics, Central China Normal University, Wuhan 430079, China}
\affiliation{$^{2}$ Artificial Intelligence and Computational Physics Research Center, Central China Normal University, Wuhan 430079, China}

\date{\today}%

\begin{abstract}
This study investigates the impact of nucleon-nucleon correlations on heavy-ion collisions using the hadronic transport model SMASH in $\sqrt{s_{\rm NN}}=3$ GeV $^{197}\rm Au$+$^{197}\rm Au$ collisions. We developed an innovative Monte Carlo sampling method that incorporates both single-nucleon distributions and nucleon-nucleon correlations. By comparing three initial nuclear configurations – a standard Woods-Saxon distribution (un-corr), hard-sphere repulsion (step corr), and ab initio nucleon-nucleon correlations (nn-corr) – we revealed minimal differences in traditional observables except for ultra-central collisions. When distinguishing between un-corr and nn-corr configurations, conventional attention-based point cloud networks and multi-event mixing classifiers failed (accuracy ~50\%). To resolve this, we developed a novel deep learning architecture integrating multi-event statistics and high-dimensional latent space feature correlations, achieving 60\% overall classification accuracy, which improved to 70\% for central collisions. This method enables the extraction of subtle nuclear structure signals through statistical analysis in high-dimensional latent space, offering a new paradigm for studying initial-state nuclear properties and quark-gluon plasma characteristics in heavy-ion collisions. It overcomes the limitations of traditional single-event analysis in detecting subtle initial-state differences.

\end{abstract}
\maketitle

\sect{Introduction}

High-energy heavy-ion collisions offer a unique opportunity to probe the structure of atomic nuclei \cite{Zhang:2017xda,Li:2008gp,Shi:2021far,Albertsson:2018maf,Pang:2019aqb,Jia:2021tzt,Giacalone:2021udy,Xu:2022ikx,Liu:2022kvz,Xu:2023ges,He:2023zin,Zhou:2023pti,Lu:2023fqd,Lacey:2010yg,Dimri:2023wup,Nie:2022gbg,Jia:2022qgl}. The much shorter time scale of these reactions ($10^{-24}$s) compared to traditional low-energy studies ($10^{-21}$s) allows for the exploration of nuclear shapes and the nucleon distribution within the nucleus \cite{Jia:2021tzt,Giacalone:2021udy,Loizides:2016djv}. The HICs generated a large amount of data about the final state hadron cloud in momentum space, at different beam energies and collision centralities, for different colliding nucleus. Various nuclear structures, including nuclear deformation, neutron skins, and alpha clusters, have been studied using HIC. Many physical observables in HICs are shown to be sensitive to the initial state nuclear structure \cite{STAR:2024wgy,Jia:2022ozr}. For instance, the centrality dependencies of charge multiplicity are sensitive to the nuclear shape deformation $\beta_2$  \cite{Gustafsson:1984ka,STAR:2015mki,Voloshin:2010ut} and the $\alpha$ cluster structure \cite{Ding:2023ibq}. The final state anisotropic flows are found to be important to determine the nuclear shape deformations $\beta_2$, $\beta_3$ and $\beta_4$ through $\langle v_n^2 \rangle \propto a_n' + b_n' \beta_n^2$ \cite{Jia:2021tzt,Giacalone:2021udy}. 
The correlation between multiplicity and anisotropic flows are found to be sensitive to the absolute values of nuclear shape deformations $|\beta_2|$ and $|\beta_4|$ \cite{Pang:2019aqb}.
The correlations between mean transverse momentum and anisotropic flows $\langle v_n^2 \langle p_T \rangle \rangle$ is demonstrated to be sensitive to the sign of nuclear shape deformation \cite{Jia:2021qyu}.

In addition to nuclear deformation, heavy ion collisions have also been used to study other nuclear structures such as neutron skins and alpha clusters. 
In the past, measurements of neutron distributions were typically made using antiproton annihilation experiments or electron parity violating scattering experiments \cite{Trzcinska:2001sy,PREX:2021umo,CREX:2022kgg}. 
Using isobaric collisions \cite{Xu:2022ikx,Liu:2022kvz,Xu:2023ges} and the ultra-peripheral collisions \cite{FANG:102102,Alvensleben:1970uw,STAR:2022wfe,Liu:2023pav}, it is also possible to determine the neutron skin thickness. Furthermore, it is also possible to extract alpha cluster information in $^{12}$C/$^{16}$O nuclei in high-energy nuclear collisions \cite{He:2021uko,Liu:2023gun,Li:2020vrg}. 
In reference \cite{He:2021uko}, the results show that by using two-dimensional histograms of the azimuthal angle and transverse momentum of the final state particles from multiple events as inputs, it is possible to accurately classify the cluster information of $^{12}$C/$^{16}$O nuclei.

Recent high-energy electron scattering experiments have revealed the existence of nucleon-nucleon short-range correlations in atomic nuclei \cite{Hen:2016kwk,JeffersonLabHallA:2007lly,Frankfurt:2008zv,Higinbotham:2009zz,Alvioli:2013qyz,Feldmeier:2011qy,Weinstein:2010rt,Malace:2014uea,Piasetzky:2006ai,Ma:2006ck}. These correlations involve pairs of nucleons with small center-of-mass momenta and large relative momenta \cite{Hen:2016kwk,Fomin:2011ng}. About $20\%$ of nucleons in atomic nuclei are involved in nucleon-nucleon pairs, mainly consisting of neutron-proton pairs\cite{Subedi:2008zz}, but they contribute significantly ($70\%$) to the total nucleon momentum \cite{Colle:2015ena,CiofidegliAtti:1995qe,Higinbotham:2014xna,Frankfurt:2008zv,Arrington:2011xs,Piasetzky:2006ai,Alvioli:2012qa}. 
The references \cite{Hen:2016kwk,Hen:2014lia} indicate that within the high-momentum $( k > k_F )$, the momentum distribution of nucleons in heavy nuclei is proportional to that of the deuteron, and it decreases with $k^{-4}$.

The relationship between the two-nucleon distribution function, the EMC effect, and short-range correlations (SRC) is remarkably tight. The ratios of the two-nucleon distribution function, denoted as $2 \rho_{2,1}(A, r) / A \rho_{2,1}(2, r)$, have been calculated with respect to the separation distance $r$ for $^3{\rm He}$ and $^4{\rm He}$ using the GFMC method alongside the AV18+UIX nucleon-nucleon interaction potential. Similarly, computations were performed for $^9{\rm Be}$ and $^{12}{\rm C}$ utilizing the VMC method with the AV18+UX potential, as referenced in \cite{Chen:2016bde}. The results obtained exhibit a good concordance with the SRC scaling factor $a_2$, which has been experimentally measured at Jefferson Lab \cite{Frankfurt:1993sp,CLAS:2003eih,Fomin:2011ng}. An intriguing linear correlation has been observed between this scaling factor and the EMC slope measured by the European Muon Collaboration \cite{EuropeanMuon:1983wih,Weinstein:2010rt,Hen:2012fm}. Notably, the scaling factor is found to be 5 to 5.5 times greater in nuclei with $A\ge 12$ than in deuterium. Furthermore, scaling factors between ${}^{4}\mathrm{He}$, ${}^{12}\mathrm{C}$, and ${}^{56}\mathrm{Fe}$ relative to ${}^{3}\mathrm{He}$ have also been measured, revealing two distinct plateaus \cite{CLAS:2005ola}. These observations can be adequately explained by theoretical models that incorporate both two-nucleon SRC and three-nucleon SRC \cite{Wang:2014rxa,Chen:2016bde}.

In Monte Carlo simulations of high energy heavy ion collisions, complex hybrid models such as color glass condensate, relativistic hydrodynamic model and hadronic transport are constructed. However, the initial configurations of nucleus are usually quite simple. For instance, spherical Woods-Saxon distribution is used in the very beginning to model the heavy ion collisions. Later on, fluctuations of nucleons inside nucleus, the nuclear shape deformations are considered at initial state. However, these are all single nucleon distributions, the effect of nucleon-nucleon correlation on heavy ion collisions have not been fully explored \cite{Baym:1995cz,Alvioli:2012pu,Blaizot:2014wba,Broniowski:2010jd,Dalal:2022zkg,Giacalone:2023hwk,Alvioli:2007jd,Piarulli:2022ulk,Chen:2003wp}. Ref.~\cite{Alvioli:2009ab} takes the nucleon-nucleon correlation into account in the Glauber model, and observed that the nucleon-nucleon correlations significantly affect the fluctuations of average number of collisions. Considering nucleon-nucleon correlations in relativistic hydrodynamic simulations of heavy ion collisions reduces the difference between elliptic and triangular flow coefficients, which improves the agreement between hydrodynamic models and the experimental measurements at LHC \cite{Denicol:2014ywa,CMS:2013bza,ALICE:2011ab}, partially resolved the ultra-central puzzle in $^{208}{\rm Pb}+^{208}{\rm Pb}$ $\sqrt{s_{\rm NN}}=2.76$ TeV collisions.

Deep learning is the state-of-the-art pattern recognition tool \cite{schmidhuber2015deep,lecun2015deep,Boehnlein:2021eym}, that has been widely used in heavy ion collisions to solve the inverse problem \cite{schmidhuber2015deep,lecun2015deep,Boehnlein:2021eym,Denby:1987rk,Larkoski:2017jix,Baldi:2016fzo,Du:2020pmp}, to extract the information of nuclear structure\cite{Pang:2016vdc}, the QCD equation of state \cite{Pang:2019aqb,Du:2019civ}, the CME signals \cite{He:2023zin,Ma:2023zfj,Wang:2023kcg,Zhou:2023pti,He:2023urp,Wang:2020tgb}. In case the traditional observables fail to identify the nucleon-nucleon correlation, we will also benchmark the performance of deep neural networks in mapping between the final state hadron distribution in heavy ion collisions and the initial state nuclear structure \cite{Bedaque:2021bja,Carleo:2019ptp,Pang:2019aqb,Epelbaum:2008ga,He:2023zin,He:2021uko,Saha:2022skj,Bhalerao:2019uzw}.

We study three types of nucleon-nucleon correlations in $^{197}{\rm Au}$ nuclei, namely un-correlation (un-corr), where the nucleons are treated as independent particles, step correlation (step corr), where the nucleons are subjected to hard sphere repulsion, and nucleon-nucleon correlation (nn-corr), which is obtained by fitting ab-initio calculation. 
We used the \trento\ initial state model to generate the initial state nuclear entropy density distribution at a centre-of-mass energy of $\sqrt{S_{\rm NN}}=2.76$ TeV with different nucleon-nucleon correlations. In addition, we utilised the SMASH relativistic molecular dynamics simulation program \cite{Weil:2016zrk}, developed at the Frankfurt Institute for Advanced Studies, to produce the final state hadron distribution at a centre-of-mass energy of $\sqrt{S_{\rm NN}}=3$ GeV.We then analysed the data to identify the effect of nucleon-nucleon correlations on the observables in heavy ion collisions. Finally, we attempted to use a deep neural network to extract the types of nucleon-nucleon correlations of the initial state nuclear structure from the final state hadron distribution in the nuclear collision.

\sect{Method}

\subsection{Sample nucleons inside nucleus with nucleon-nucleon correlations}

To generate the initial condition of heavy ion collisions for relativistic hydrodynamic simulations and transport model studies, 
Woods-Saxon distribution is widely employed to sample the coordinates of nucleons inside nucleus. 
For spherical nucleus without shape deformation, the radial coordinate of each nucleon with respect to the center of nucleus is simply given by,
\begin{align}
    \rho(r) = \frac{\rho_0}{\exp(\frac{r-r_0}{d})+1},\\
    A = \int_0^{\infty} 4\pi r^2 \rho(r) dr,
    \label{eq:woodsaxon}
\end{align}
where $A$ is the number of nucleons in the nucleus, $\rho_0$ is a normalization factor, $r_0\approx 1.25R^{1/3}$ is the nuclear radius,
$d$ is the surface thickness of the nucleus.

However, sampling the coordinates of nucleons following the Woods-Saxon distribution suffer from a few problems.
The distribution is the single nucleon distribution, nucleons sampled from this distribution are independent of each other.
As a result, the sampled nucleons have larger fluctuations than in the realistic case.
For example, two or more nucleons may overlap, whose probability might be suppressed by the strong short range repulsion core
in the nucleon-nucleon interaction potential.
On the other hand, holes may form among several sampled nucleons that is also disfavored by the long range attractive force between nucleons.
To produce a more realistic initial condition for heavy ion collision, 
one should further take into account the relative distance distribution of two nucleons $\rho(\Delta r)$,
where $\Delta r = |\mathbf{r_1}-\mathbf{r_2}|$.
To get $\rho(\Delta r)$, the two-nucleon distribution function $\rho(\mathbf{r_1},\mathbf{r_2})$ can be reformed as
$\rho(\mathbf{R},\mathbf{\Delta r})$, and 
\begin{equation}
    \rho(\Delta r)=\int d\mathbf{R}\rho(\mathbf{R},
    \mathbf{\Delta r}),
\end{equation}
where $\mathbf{R}={1\over 2}(\mathbf{r_1}+\mathbf{r_2})$ is the mass center of each pair of nucleons.

The probability two-nucleon density distribution of can be approximated by,
\begin{equation}
    \rho(\mathbf{r_1},\mathbf{r_2}) \approx \rho(r_1)\rho(r_2)[1-C(\Delta r)],
\end{equation}
with 
\begin{equation}
    C(\Delta r) = 1 - \rho_c(\Delta r)/\rho_u(\Delta r).
    \label{c_deltar}
\end{equation}
Here $\rho_c(\Delta r)$ and $\rho_u(\Delta r)$ represent the relative distance distribution of nucleon pairs with and without consideration of two-nucleon distribution functions.

Ref~\cite{TabatabaeeMehr:2024lgu} reviewed four distinct sampling methodologies for two-nucleon correlations with minimum distance constrains, $|\vec{r}_i - \vec{r}_j| \ge d_{\rm min}$:
\begin{itemize}
    \item The Direct Method generates nucleon positions according to the Woods-Saxon (WS) distribution and rejects any sampled nucleus if at least one nucleon pair violates the minimum distance criterion. While straightforward to implement, this approach suffers from low computational efficiency.
    \item The Iterative Method \cite{moreland2015alternative} pre-generates radial coordinates based on WS distributions and iteratively samples and adjusts angular configurations to eliminate nucleon-nucleon distance violations.
    \item The Shifting Method in Ref~\cite{Luzum:2023gwy,Luzum:2023oic} 
slightly displaces $\vec{r}_i$ and $\vec{r}_j$ along their relative position vector in opposite directions with respect to their center, ensuring compliance with the nucleon-nucleon correlation function. However, this method introduces minor deviations in the single-nucleon distribution.
\item The Markov Chain Monte Carlo (MCMC) Method \cite{Alvioli:2009ab} employs random walks in nuclear configuration space to sample nucleon positions.
\end{itemize}

In this work, we adopt the Iterative Method, which ensures that sampled nucleons satisfy both the single-nucleon distribution and nucleon-nucleon correlation functions.

\begin{figure}[htp]
\centering
\includegraphics[width=0.48\textwidth]{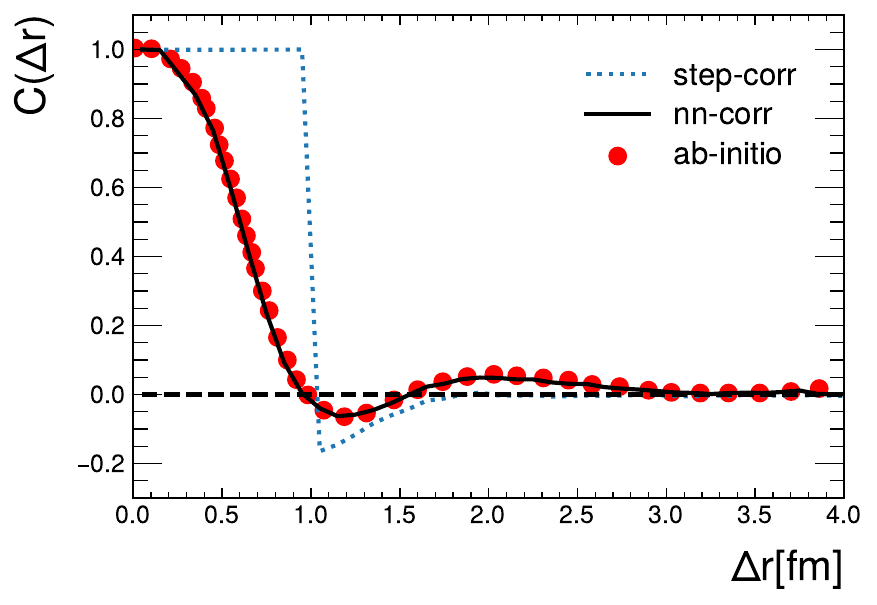}
\caption {\label{fig:c_dr}(Color online) The nucleon-nucleon correlations of un-corr, step corr and nn-corr according to the correlation function.The blue dashed and black solid lines correspond to the step corr and nn-corr types respectively, while the red dotted line is from the ab-initial calculation in \cite{Alvioli:2005cz,Alvioli:2009ab}.
}
\end{figure}

Fig~\ref{fig:c_dr} illustrates the correlation function of sampled nucleons under various scenarios: the independent nucleons (un-corr) represented by a black dashed line, the hard sphere repulsion (step corr) depicted by a blue dotted line, and the ab-initio nucleon-nucleon correlation (nn-corr). The black solid line denotes the sampled nucleons, while the red circles correspond to theoretical computations.

(1){\it un-corr type} corresponds to initial conditions without nucleon-nucleon correlation. 
Since the shape of the $^{197}Au$ nucleus is approximately spherical, the radial distribution of nuclei within it can be described as a Wood-Saxon distribution as given by Eq.~\ref{eq:woodsaxon}. For gold nucleus, parameters are chosen to be $\rho_0=0.168$ and $r_0=6.38$ and $d=0.535$. The probability distribution of $A$ nucleons in spherical coordinates reads,
\begin{equation}
\begin{aligned}
    P(r,\theta,\phi)&=p(r)p(\theta)p(\phi)drd\theta d\phi   \\
    &={{\rho(r)r^2sin\theta drd\theta d\phi}\over{A}},
\end{aligned}
\end{equation}
where $\rho(r)$ is the Woods-Saxon distribution function. Thus the probability density functions for the radial coordinate $r$, the polar angle $\theta$ and the azimuthal angle $\phi$ are given by
\begin{equation}
\begin{aligned}
    p(r) &= {4\pi r^2 \rho(r) / A} \ ,\ r \in [0, \infty], \\
    p(\theta) &= {\sin\theta \over 2}\ ,\ \theta \in [0, \pi], \\
    p(\phi) &= {1 \over 2\pi}\ , \ \phi \in [0, 2\pi].
\end{aligned}
\end{equation}

We use rejection sampling to sample the radial coordinate $r \sim P(r)$, and direct sampling for the polar angle $\theta \sim P(\theta)$ and the azimuthal angle $\phi \sim P(\phi)$, respectively.

(2){\it step corr type} stands for hard sphere repulsion for the two-nucleon distribution functions. In this case, the nucleon coordinates are obtained by setting the minimum nucleon-nucleon distance "$d_{min}$" in the \trento\ model. Here we choose $d_{min}=1$ fm. The \trento\ model initially employs the Wood-Saxon distribution to sample the radial distances of nucleons from the center, followed by the random sampling of polar and azimuthal angles. Subsequently, the Cartesian coordinates of the nucleons are computed using the standard formula for converting spherical coordinates to Cartesian coordinates.

Then the relative distance $\Delta r_i$ between the newly sampled nucleon and all previously accepted nucleons is calculated, where $(x,y,z)$ is the new nucleon and $(x_i,y_i,z_i)$ are the accepted nucleons in the Cartesian coordinates. If the new nucleon is too close to one of the accepted nucleons (i.e. $min(\Delta r_i)<1$ fm), the polar and azimuthal angles of the nucleon are re-sampled to obtain a new coordinate $(x',y',z')$. This procedure does not affect the radial distribution of the nucleons. If the number of resamples exceeds a certain threshold, the last sampled nucleon is accepted. For the minimum nucleon-nucleon distance $d=1$ fm, the number of such nucleons is relatively small compared to the total number of nucleons and has a negligible effect on the sampling result.

\begin{table*}[!htb]
\begin{center}
\caption{Coefficients of polynomial function $C(\Delta r)=\sum_{i=0}^{n}a_i \Delta r^i$.}
\begin{tabular}{|c|c|c|c|c|}
    \hline
    $a_0$-$a_3$   &1.00547756316    &-0.09118891837        &2.28906649136     &-34.11966099655\\
    \hline
    $a_4$-$a_7$    &204.85019736951    &-735.87975376369        &1619.73061995975     &-2325.40229004033\\
    \hline
    $a_8$-$a_{11}$   &2292.23465815435    &-1596.85102844322        &789.97353268919     &-269.16568289196\\
    \hline
    $a_{12}$-$a_{15}$   &55.53686919901    &-2.26452740059        &-2.69447194442     &1.00172060913\\
    \hline
    $a_{16}$-$a_{19}$  &-0.19068811248    &0.02181921704        &-0.00142837921     &0.00004143056\\
    \hline
\end{tabular}
\label{table:coefficients}
\end{center}
\end{table*}

(3){\it nn-corr type} corresponds to ab initio nucleon-nucleon correlation. 
By polynomial fitting, we obtain an approximate expression for the correlation function $C(\Delta r)$ in the interval $\Delta r \in [0,4]$ fm:
\begin{align}
    C(\Delta r)=\sum_{i=0}^{n}a_i \Delta r^i,
\end{align}
where $a_i$ represents the coefficient of the i-th order whose values are given in Table.~\ref{table:coefficients}. Using formula \ref{c_deltar}, we can calculate the second term on the right-hand side:
\begin{equation}
    P_{accept}(\Delta r)= 1 - C(\Delta r),
    \label{P_deltar}
\end{equation}
which is the acceptance probability of nucleons with relative distances in the interval $\Delta r\in [0,4]$ fm.

First, we sample the radial coordinates of 197 nucleons in a $^{197}Au$ nucleus from the Woods-Saxon distribution. Then, we sequentially generate the polar and azimuthal angular coordinates for each nucleon, while calculating the relative distances between the newly sampled nucleon and all previously positioned nucleons.
The probability of accepting this nucleon is $\prod_i [1 - C(\Delta r_i)]$, where $\Delta r_i$ is the relative distance between the coordinates of the newly proposed nucleon and the $i$-th previously accepted nucleon.
The distribution of relative distance between the newly sampled nucleon and the accepted nucleons should match the ab-initio calculations as shown in Fig.~\ref{fig:c_dr}. If the new coordinate is rejected, the above process is repeated until  it is accepted.

\subsection{Initial state of high energy heavy ion collisions from \trento}

\trento\ is a parameterised model for simulating the initial states of high-energy nucleus-nucleus collisions \cite{moreland2015alternative}. It takes input information such as the initial state of the colliding nuclei, the minimum distance between the nucleons and the nucleon-nucleon collision cross section to generate a realistic initial entropy density distribution and an anisotropic distribution of the collision initial states in the transverse plane.

In \trento, a pair of nuclei A and B collide inelastically along the z-axis. Since nuclear-nuclear collisions can be viewed as a superposition of nucleon-nucleon collisions, we can discard the nucleons that did not collide by sampling the collision probability, and the remaining nucleons are the participants in the collision process. 

The scalar field $f(T_A,T_B)$ is defined to convert the thickness fluctuation of fast-moving nuclei into entropy deposition, which is proportional to the entropy generated in the central rapidity region during the fluid hydrodynamics thermalization time $f\propto dS/dy|_{r=r_0}$. 
Therefore, by calculating the reduced thickness $T_R$, we can obtain the anisotropy of the initial entropy density distribution in the transverse plane:
\begin{equation}
\begin{aligned}
&dS/dy|_{r=r_0}\propto T_R(p;T_A,T_B),\\
&f=T_R(p;T_A,T_B)=\left({{T_A^p+T_B^p}\over{2}}\right)^{1/p},\\
&T_{A,B}(x,y)=\int dz \rho_{A,B}(x,y,z),
\end{aligned}
\end{equation}
where the reduced thickness parameter $p$ can change the influence of the sampled nuclear thickness on the entropy density. 
The spatial distribution of the nucleons is transformed into the transverse plane by Lorentz-boosting, thereby losing part of the information about the spatial coordinate distribution of the nucleon, and finally converted into entropy deposition in mutual collisions.

The eccentricity of a single event is defined as $\epsilon_n$,
\begin{align}
    \epsilon_n e^{in\phi}=-{{\int dxdy r^n e^{in\phi}T_R}\over{\int dxdy r^nT_R}},
\end{align}
where $(r,\phi)$ represents the transverse position of participating nucleons with respect to the centre of mass.

In our research on nucleon-nucleon correlations, we have modified the minimum nucleon-nucleon distance parameter $d_{min}=1$ fm in the \trento\ model for sampling initial nuclear configurations, in order to generate un-corr and step-corr types of nuclei. Other important parameters include a collision energy of$\sqrt{S_{\rm NN}} = 2.76$ TeV, corresponding to a nucleon-nucleon scattering cross section of 6.4 fm$^2$;  a reduced thickness parameter $p=0$; and a fluctuation parameter $k=1$.

\subsection{Intermediate energy heavy-ion collisions simulated using SMASH}

SMASH is mainly used to simulate the resonance scattering and decay of final state hadrons in high-energy nuclear collisions, as well as the overall dynamics of low-energy nuclear collisions. The SMASH simulation programme solves the relativistic Boltzmann equation using molecular dynamics:
\begin{equation}
p^\mu \partial_\mu f_i(x,p)+m_iF^\alpha \partial^p_\alpha f_i(x,p)=C^i_{coll},
\end{equation}
where $C^i_{coll}$ is the collision term, $f_i$ is the single-particle distribution function, $m_i$ and $p^\mu$ are the mass and four-momentum of the particle, and $F$ is any additional force acting on the particle. 

In SMASH, the criterion for collision between two particles is that the transverse distance between particle a and particle b is less than the interaction distance $d_{ab}<d_{int}=\sqrt{{\sigma_{total}/{\pi}}}$, where $\sigma_{total}$ is the total scattering cross section between the two particles and $d_{ab}$ is the transverse distance between a and b in the relativistic case:
\begin{equation}
d_{ab}^2=(\mathbf{r_a}-\mathbf{r_b})^2-{{(\mathbf{r_a}-\mathbf{r_b})\cdot (\mathbf{p_a}-\mathbf{p_b})^2\over{(\mathbf{p_a}-\mathbf{p_b})^2}}}.
\end{equation}
Here, $\mathbf{r}$ and $\mathbf{p}$ are the coordinates and momenta in the centre of mass frame.

In this article, we study nucleon-nucleon correlations by using the SMASH collision model to simulate collisions between nuclei of the same type that are read in from an external source.
Smash requires the Cartesian coordinates and isospin of each nucleon for every initial-state gold nucleus. There are two sets of data files provided, representing projectile and target nuclei respectively, are provided as input to Smash. Each file contains 600,000 gold nuclei, with the first 79 nucleons representing protons and the next 118 representing neutrons for each gold nucleus.

Before the collision, the nucleus undergoes a random Euler angle rotation, and the nucleons in the nucleus acquire new coordinates. During the collision, the nucleons in the nucleus undergo free flight, elastic collision, resonance decay and production processes. The transverse momentum of the nucleons before the collision is $(p_x, p_y) = (0, 0)$ GeV/c, and the longitudinal momentum $p_z$ depends on the collision energy $\sqrt{S_{\rm NN}}$, with $E=\sqrt{S_{\rm NN}}/2$ GeV. In this study we mark the nucleons with zero transverse momentum after the collision as spectators and exclude them from the data analysis. To improve the robustness of the model, we randomly reorder the particles in each event. 

We investigated the influence of two-nucleon distribution functions on final-state particle momentum distributions using the SMASH hadronic transport model at a center-of-mass collision energy of $\sqrt{S_{\rm NN}}=3$ GeV. 
The study employed both un-corr and nn-corr configurations of the $^{197}\rm Au$ nucleus within the SMASH framework incorporating mean field potential, generating 500,000 collision event samples for analysis.

\subsection{Multi-event correlations of latent features by deep neural network}

\begin{figure*}[htp]
\centering
\includegraphics[width=1\textwidth]{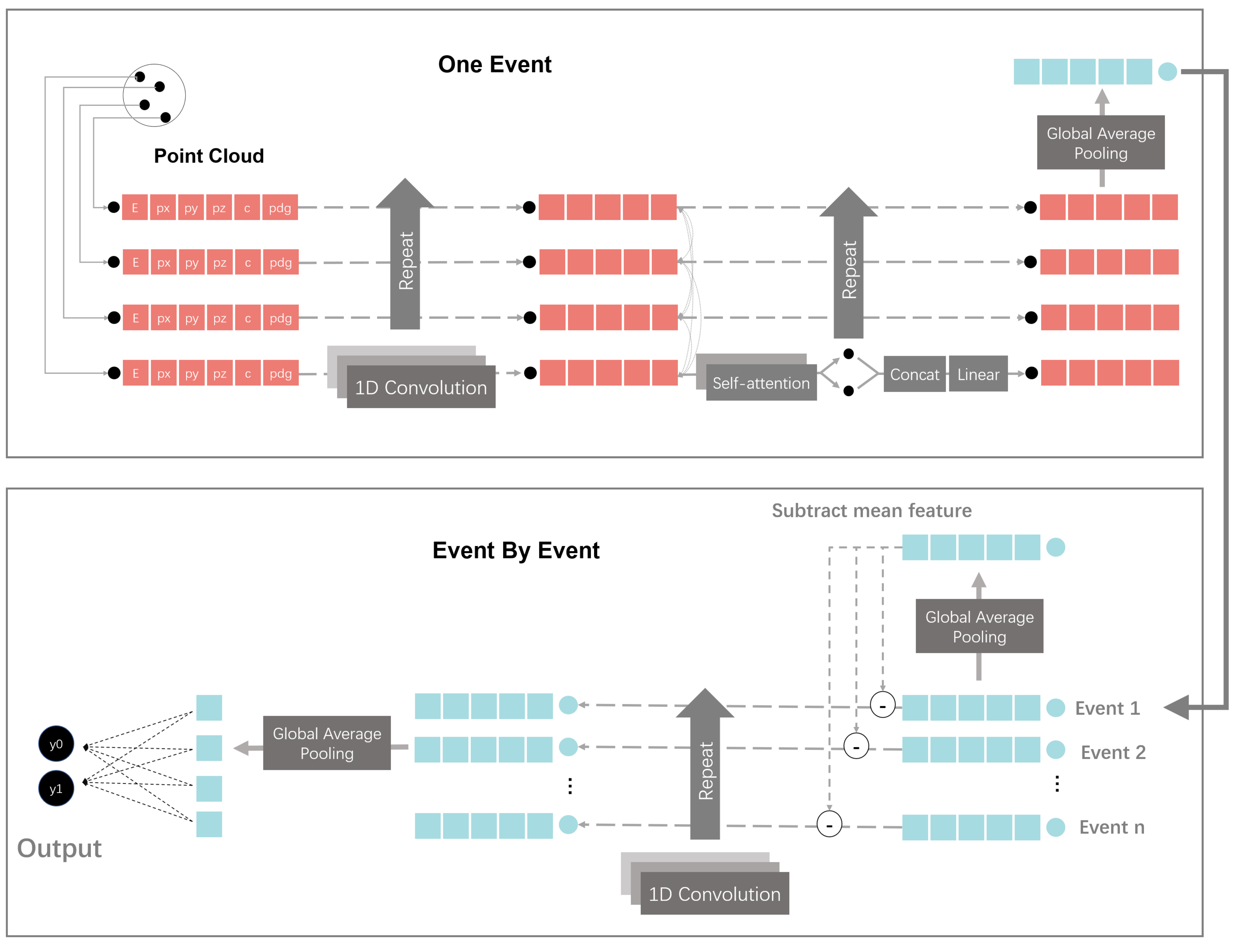}
\caption {\label{fig:network}(Color online) The neural network architecture. The top diagram shows the use of a point cloud neural network with a multi-head attention mechanism to learn the correlations between particles within an event, ultimately outputting a feature vector representing a single event. The lower diagram illustrates the process of learning statistical features across multiple events within a sample.
}
\end{figure*}
Point cloud data is a collection of points in a particular coordinate system that exhibit spatial transformation invariance and correlation, providing rich information including 3D coordinates (X, Y, Z), time, color and intensity. We have found that this type of neural network architecture is naturally suited to analysing data generated from particle collision experiments. This kind of data contains a large amount of momentum space coordinates and various quantum number information of final state hadrons. If we convert this final state hadron information into images and input them into traditional convolutional neural networks for analysis, not only will the hadron momentum information become sparse in a 2D distribution map, but the accuracy of the momentum and the quantum number information of the hadrons will also be lost. Point cloud neural networks can directly process list-type data consisting of multiple particles, each of which carries various attributes such as four momentum, mass and charge.

The one-dimensional convolutional layers are the main part of the point cloud neural network. It can convert the features of multiple particles into a representation in a latent variable space where all particles share the same convolution kernel. In this process, each particle in the same event is transformed individually into the high-dimensional latent variable space, but the correlation between particles is not taken into account. We tried to use a network architecture of (n, 128×3) to extract the high-dimensional information of particles, where n represents the features of each particle, and this architecture converts n features of particles into 128 features of latent variable space, and the hidden layer used to extract features has 3 layers.

We hope that neural networks can learn the statistical information between events. The conventional method is to concatenate the particle information of several events into one event and input it into the neural network for analysis. However, this approach may lose the particle number information of individual events, and the particles in different events should be independent of each other.

Therefore, after converting the momentum information of the particles into high-dimensional latent variables, we set the parameter axis in the two multi-head Attention layers \cite{vaswani2017attention} to learn the correlation between particles within an event and the correlation between events. The latter represents the statistical information of multiple events.

To improve the learning ability of the model, we put the output of the multi-head self-attention layer back into the one-dimensional convolutional network with a (128, 128×1) architecture. Then, we use the global maximum pooling technique in deep learning to calculate the maximum value of the high-dimensional latent variables of all particles in each dimension, that is, to convert the 128-dimensional latent variables into 128 floating-point numbers, each of which represents the long-range correlation between particle-particle in each event.

Next, we send the maximum values of the 128 latent variables into a fully connected network. For the binary classification task, the output of the network has 2 neurons representing the predicted probability of two different things. Note that the output layer uses the softmax activation function, while other layers use the relu activation function. To ensure that the neural network has good convergence and generalisation in function space during the learning process, we added batch normalisation layers after each layer of the network.

In this study, we used a network architecture consisting of a point cloud network and a multi-head self-attention mechanism to take the final state hadron information in events as input to the neural network, as depicted in Fig~\ref{fig:network}. 
Since particles in different events are not correlated, we input the particle features from individual events into a neural network. The input features consist of six dimensions: $(E, p_x, p_y, p_z, \text{charge}, \text{pid})$. The first four dimensions represent the four-momentum information of the hadron, charge represents the electric charge of the hadron, and pid is the hadron's coding, represented by a 5-element array. The first column of the array indicates whether the hadron is a particle or an antiparticle, represented by 0 and 1 respectively, and the remaining four columns represent the PDG code of the hadron. If the PDG code of the hadron is less than four digits, it is padded with zeros at the front. For example, a proton is represented as [0, 2, 2, 1, 2], and a $\pi^-$ particle is represented as [1, 0, 2, 1, 1]. 
We use one-dimensional convolution to map the single-particle momentum information into a high-dimensional feature space. The multi-head attention mechanism is employed to learn the pairwise and multi-particle correlation information in the high-dimensional feature space. Finally, global average pooling is used to obtain the average feature representing each individual event.

In traditional statistical methods, after obtaining a set of observables, it is a common practice to calculate the correlation between these observables.
For instance, if there are two observables $a$ and $b$, the correlations between $a$ and $b$ is given by,
$$C(a, b) = \overline{\delta a \delta b} = \overline{(a - \overline{a})(b - \overline{b})}$$
where $\delta a = a - \overline{a}$, $\overline{a}$ is the event-average value of $a$.

Similarly, to enable the neural network to learn the high-dimensional statistical information about multiple events, we use global average pooling to get the mean value of each feature among multiple events. 
These mean values are subtracted from the features of each event,
and feed back to the neural network for further processing.

We found a total of 17 types of final state hadrons in these collision events, and since the number of final state particles produced in each event varies, we chose 700 as the maximum number of particles, and events with fewer particles were filled with zero vectors.
We combine fifty events into one sample, with each sample corresponding to one label. After the neural network learns the high-dimensional statistical information across multiple events, it passes through the global average pooling layers and fully connected layers to obtain the prediction results for individual samples.

\sect{Results}
\subsection{Look for nn-corr from the nucleon distribution of sampled nuclei}

\begin{figure*}[htp]
\centering
\includegraphics[width=1\textwidth]{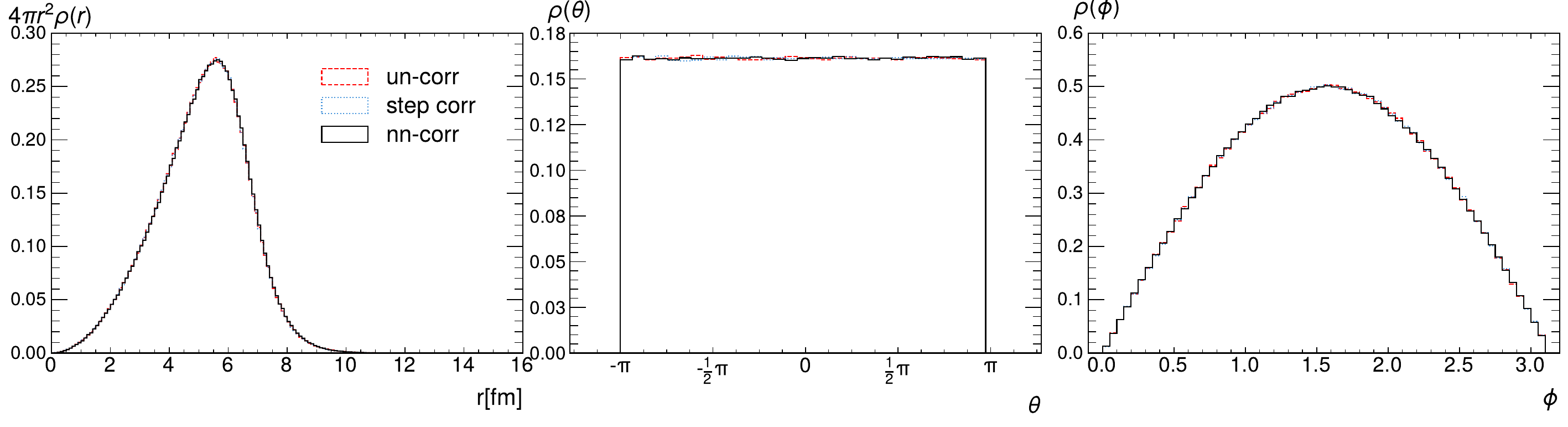}
\caption {\label{fig:rthetaphi_dist}(Color online) The sampling results of $^{197}Au$ are shown in the figure. 
The left, middle, and right figures respectively represent the radial coordinate distribution $dN/dr$, polar angle distribution $dN/d\theta$, and azimuthal angle distribution $dN/d\phi$ of nucleons for three types of the two-nucleon distributions. The radial distribution $r$ with consideration of the two-nucleon distributions should be the same as without. The blue dotted line, red dashed line, and black solid line represent the un-corr, step corr, and real nn-corr types, respectively.
}
\end{figure*}

Fig~\ref{fig:rthetaphi_dist} shows the single nucleon distribution along the $r$, $\theta$ and $\phi$ directions for sampled nuclei under three distinct nucleon-nucleon correlation scenarios: un-corr, step corr, and nn-corr. It is evident that the single nucleon distributions remain unaffected by the specific nucleon-nucleon correlation function employed during the sampling process. The $r$ distribution is straightforward, as the radial coordinates of all nucleons are sampled from the Woods-Saxon distribution. However, the $\theta$ and $\phi$ distributions present a more complex scenario, as these two coordinates for each nucleon have been adjusted to ensure that the newly sampled nucleon adheres to the prescribed nucleon-nucleon correlation constraints.

\begin{figure}[htp]
\centering
\includegraphics[width=0.48\textwidth]{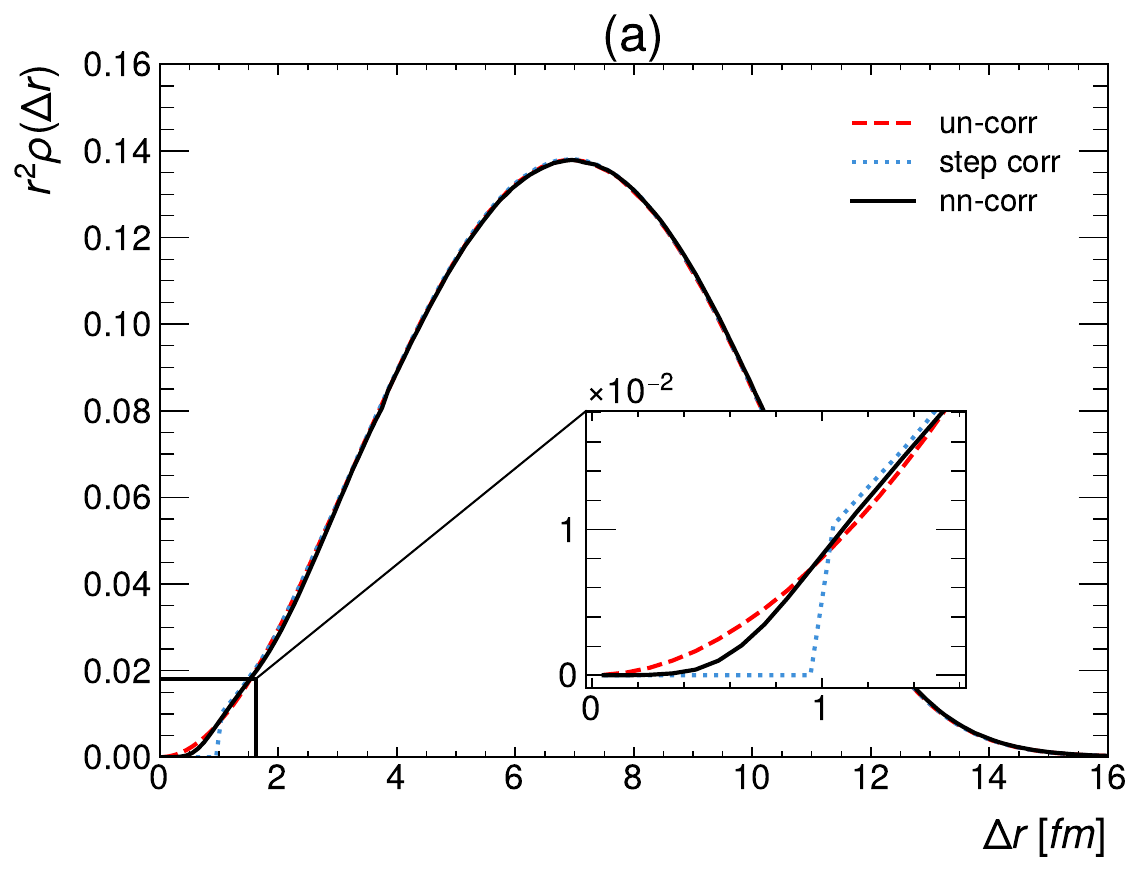}
\includegraphics[width=0.48\textwidth]{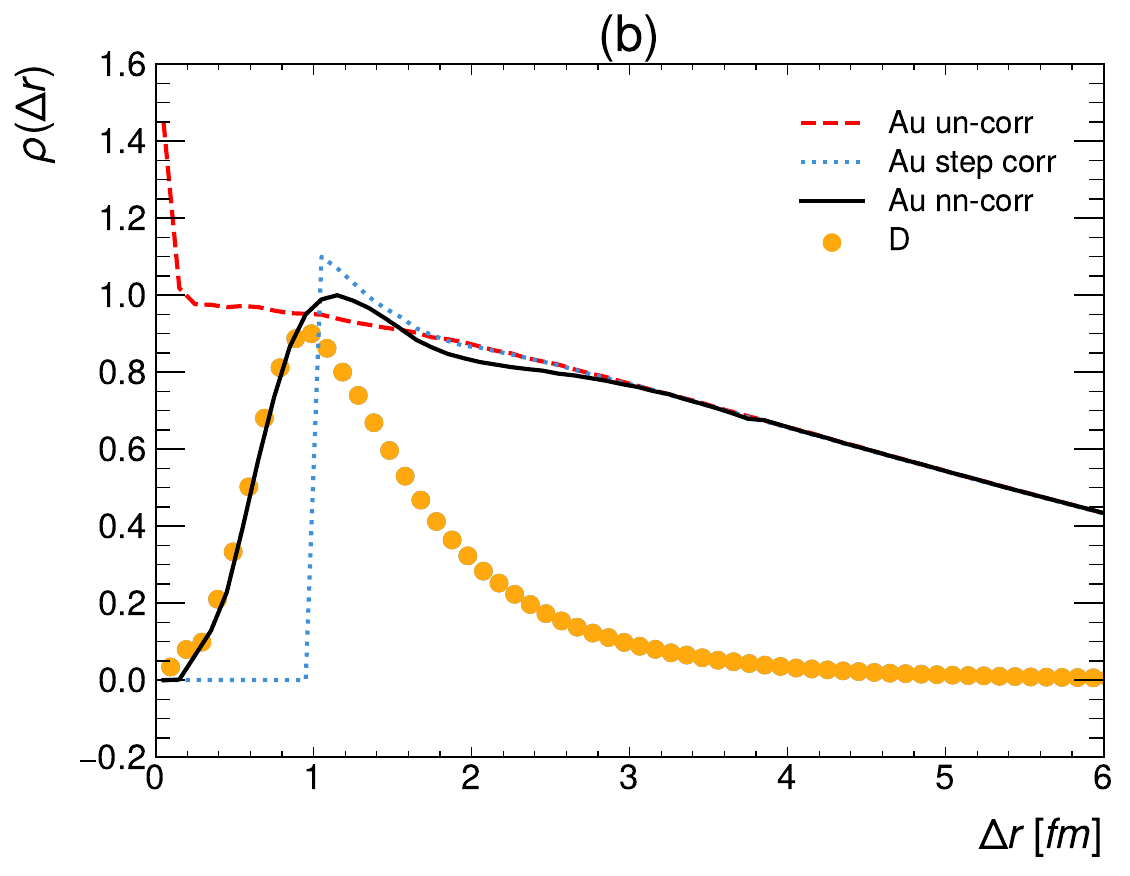}
\caption {\label{fig:nnc_initial_dist}(Color online) The two-body probability distribution $P(\Delta r)$ and two-nucleon density distribution $\rho(\Delta r)$ for three types of the nucleon-nucleon correlations.
}
\end{figure}

Fig~\ref{fig:nnc_initial_dist} shows the probability distribution of nucleon relative distances (upper panel), and the corresponding two-nucleon density distributions (lower panel) for three nucleon-nucleon types. The most pronounced discrepancies between these distributions occur in the [0,1] fm range, while at larger relative distances they show remarkable similarity.
Since universality exists for two-body densities of different nuclei at small distances and high momentums \cite{Feldmeier:2011qy}, 
we compare the two-nucleon density distributions sampled from three types of nucleon-nucleon correlations and compare them with the deuteron wave function, as illustrated in the lower panel of Fig~\ref{fig:nnc_initial_dist}. 
In the region of small relative distance ($\Delta r \leq 1$ fm), the $\rho(\Delta r)$ distribution of Au nucleus with nn-corr is in good agreement with the the $\rho(\Delta r)$ of deuterons. 
Where the deuteron wave function is given by~\cite{Zhaba:2015yxq,Zhaba:2017syr,Wiringa:1994wb}, 
\begin{align}
    \Psi_{d}=\psi_{S}+\psi_{D}=\frac{u(r)}{r} Y_{101}^{1}+\frac{w(r)}{r} Y_{121}^{1}
\end{align}
Where $u(r)$ and $w(r)$ are radial deuteron wave functions for states with the orbital momentum $l=$ 0 and 2, $y_{101}^{1}$ and $y_{121}^{1}$ are the vector spherical harmonics.

\subsection{ Look for nn-corr using the entropy density distribution of heavy ion collisions simulated by \trento}

\begin{figure}[htp]
\centering
\includegraphics[width=0.48\textwidth]{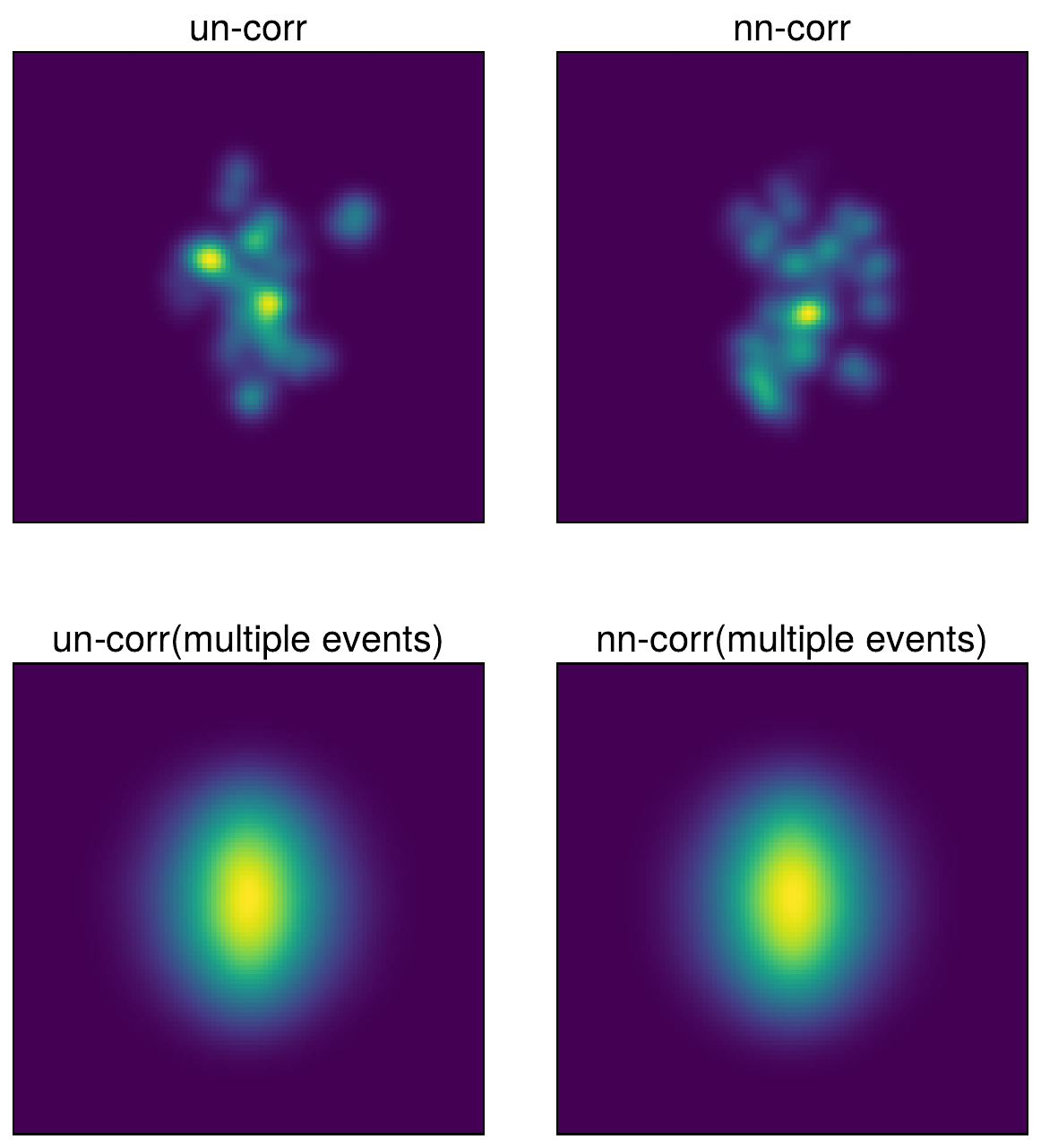}
\caption {\label{fig:trans}(Color online) The heat map of the entropy density distribution generated by \trento. The upper figures show the single event of the type un-corr and nn-corr types; the lower figures show the cumulative entropy density distribution of several events.
}
\end{figure}

Fig~\ref{fig:trans} visualizes the initial state entropy distribution in high-energy nucleus-nucleus collisions through a two-dimensional heatmap (100×100 grid) derived from \trento\ model simulations.
However, it is difficult to identify the differences of the entropy density distribution between un-corr and nn-corr, visually.

\begin{figure*}[htp]
\centering
\includegraphics[width=1\textwidth]{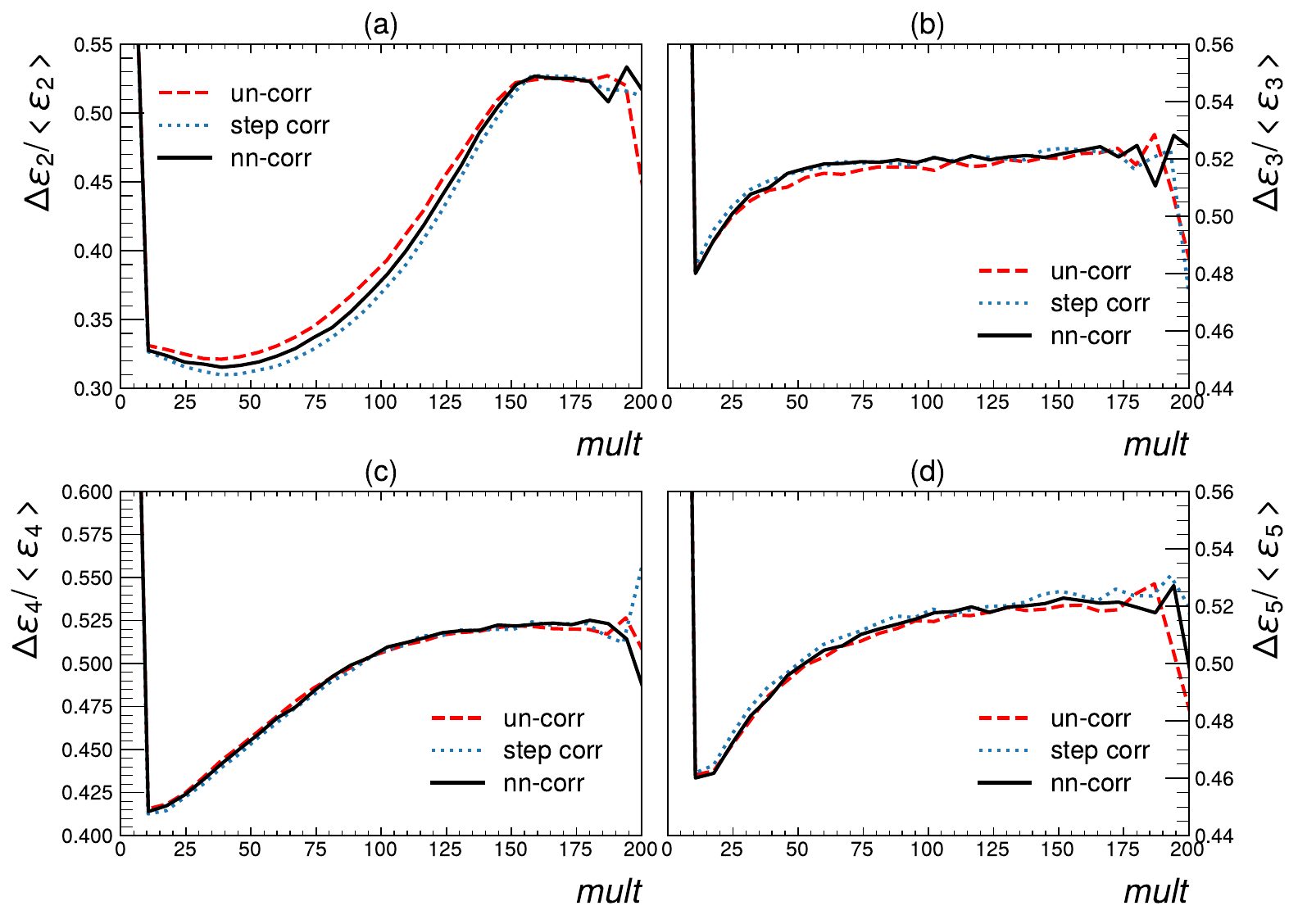}
\caption {\label{fig:var_ecc}(Color online) The correlation between the scaled standard deviation of participant eccentricity, ${\Delta \epsilon_n/<\epsilon_n>}$, and the integrated reduced thickness, ${mult}$, generated by \trento\ for three types of two-nucleon distributions. Figure~(a), (b), (c), (d) show the scaled standard deviation when n = 2, 3, 4, 5 respectively.
}
\end{figure*}

Fig~\ref{fig:var_ecc} illustrates the event-by-event fluctuation of eccentricities as a function of the integrated reduced thickness, which is proportional to the entropy generated at mid-rapidity \cite{moreland2015alternative}. 
The event-by-event fluctuation of eccentricities, the scaled standard deviation $\Delta \epsilon_n/\textless\epsilon_n \textgreater$, where $\epsilon_n$ denotes the eccentricity of a single event, is sensitive to different nucleon-nucleon correlations \cite{Broniowski:2010jd}. 
Notably, only the fluctuation of the second-order eccentricity exhibits a significant difference. In the central region, the un-corr curve is higher than the nn-corr curve, with the step corr curve being the lowest.

\begin{figure}[htp]
\centering
\includegraphics[width=0.48\textwidth]{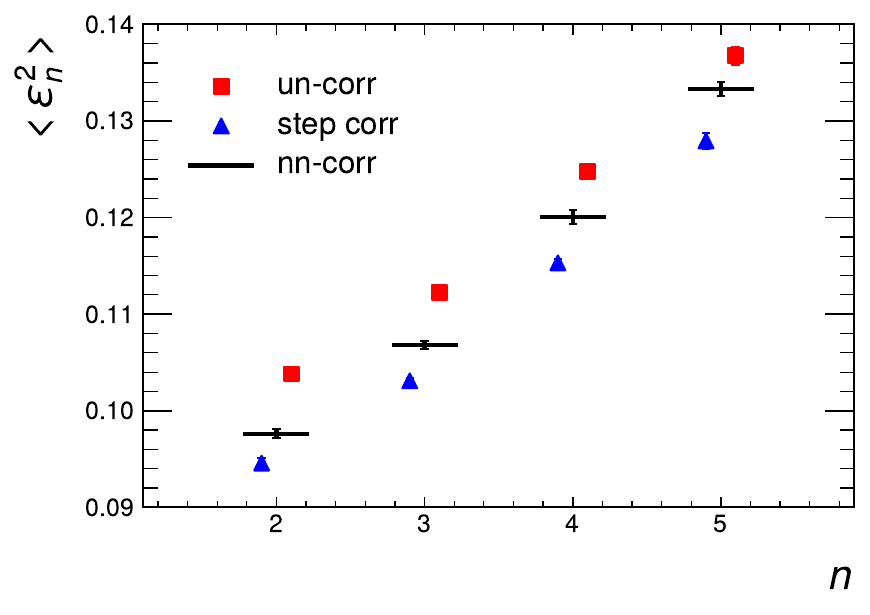}
\caption {\label{fig:e_cent_01}(Color online) Comparison of the three types of the two-nucleon distribution function $\epsilon_n\{2\}$ at $0\%$ to $1\%$ centrality. Black lines, blue triangles and red squares indicate the results for the un-corr type, step corr type and nn-corr type, respectively.
}
\end{figure}

Fig~\ref{fig:e_cent_01} shows the effect of the two-nucleon distributions on $\epsilon_n\{2\}$ in ultra-central collisions by analyzing collision events with centrality ranging from $0\%$ to $1\%$, defining the root mean square of the eccentricities $\epsilon_n\{2\}$, 
\begin{align}
    \epsilon_n\{2\}=\sqrt{\textless\epsilon_n \epsilon_n^*\textgreater_{events}},
\end{align}
where $\epsilon_n$ is the eccentricity of the single event. Both types of the two-nucleon distributions result in significantly smaller values of $\epsilon_2\{2\}$ and $\epsilon_3\{2\}$ compared to the values without nucleon correlations, which is consistent with the findings in \cite{Denicol:2014ywa}.

\subsection{Look for nn-corr in particle clouds from SMASH simulations of heavy ion collisions}

\begin{figure}[htp]
\centering
\includegraphics[width=0.48\textwidth]{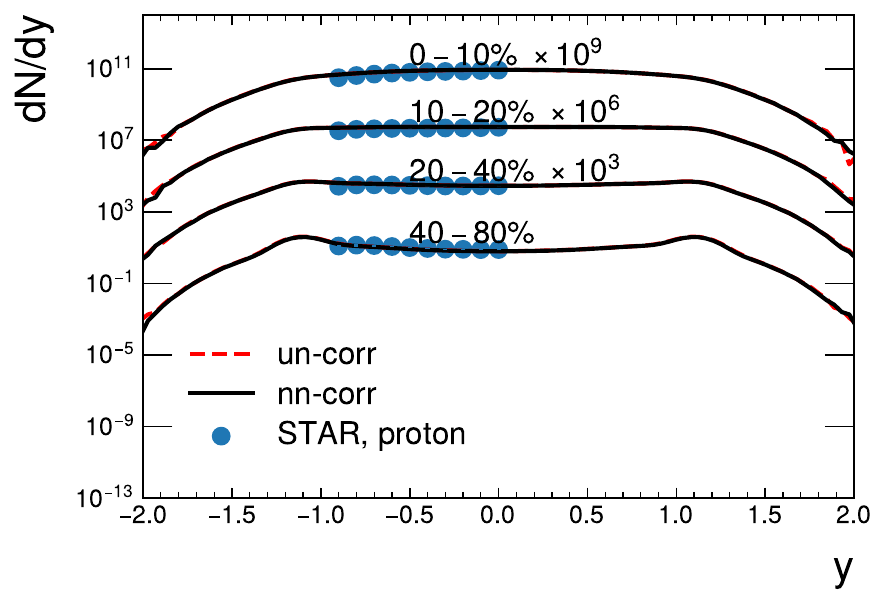}
\includegraphics[width=0.48\textwidth]{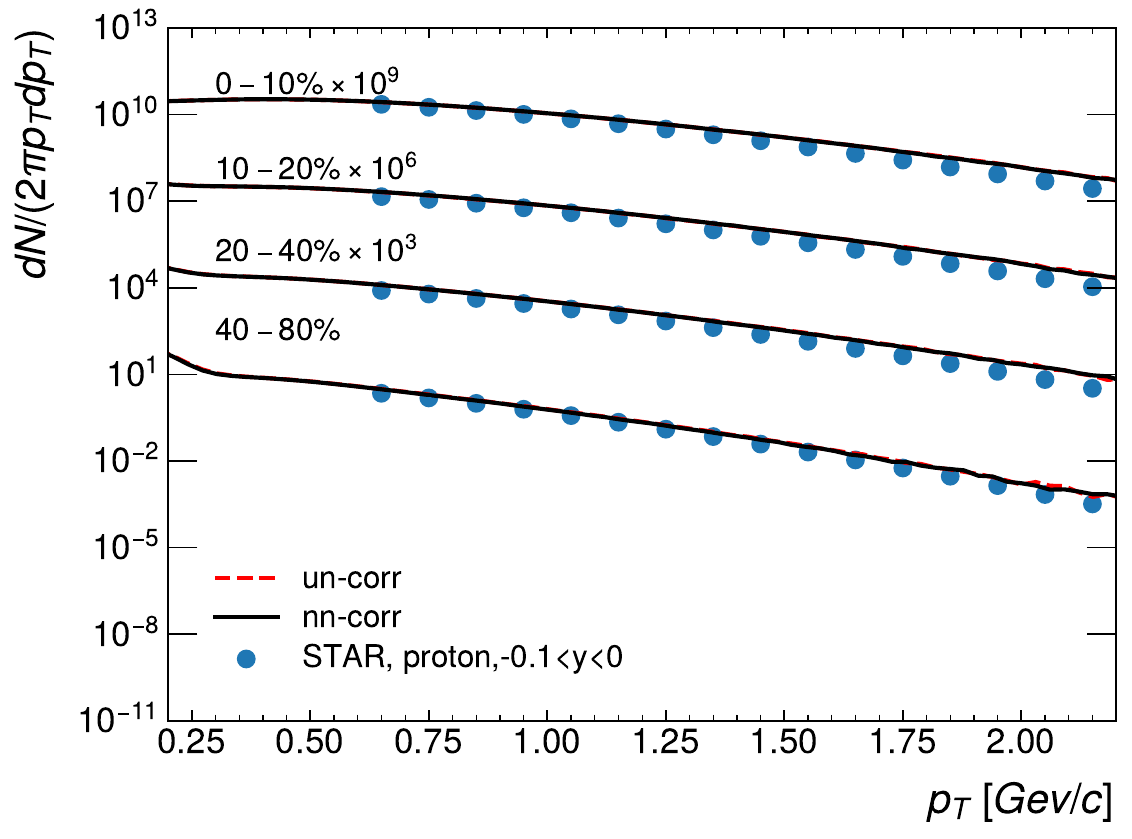}
\caption {\label{fig:nnc_ptY}(Color online) The proton distribution of the events in the final state for the un-corr and nn-corr types with different centralities. The top figure shows the event distribution for the rapidity and the bottom figure shows the event distribution for the transverse momentum.
Comparison of SMASH results with experimental data from STAR Au+Au collisions at $\sqrt{S_{\rm NN}}=3$ GeV \cite{STAR:2023uxk}.}
\end{figure}

Fig~\ref{fig:nnc_ptY} analyzes their distribution of transverse momentum and rapidity with different centralities in $^{197}{\rm Au}$+$^{197}{\rm Au}$ collisons at $\sqrt{S_{\rm NN}}=3$ GeV. 
All the events divided into 4 centrality regions: 0-10\%, 10-20\%, 20-40\%, and 40-80\% centrality regions have 50,000, 50,000, 100,000, and 200,000 events, respectively. For a comparative analysis with experimental data, we calculate the rapidity distribution and the transverse momentum distribution of the proton for $-0.1 < y < 0$, which limits the transverse momentum of the proton to $p_T > 0.2$ GeV/c. Nucleon-nucleon correlations have a small effect on the rapidity and transverse momentum distributions at $\sqrt{S_{\rm NN}} = 3$ GeV, and the transverse momentum distribution of the proton is larger than the experimental one in the large $p_T$ region due to the effect of the mean field.

\begin{figure}[htp]
\centering
\includegraphics[width=0.48\textwidth]{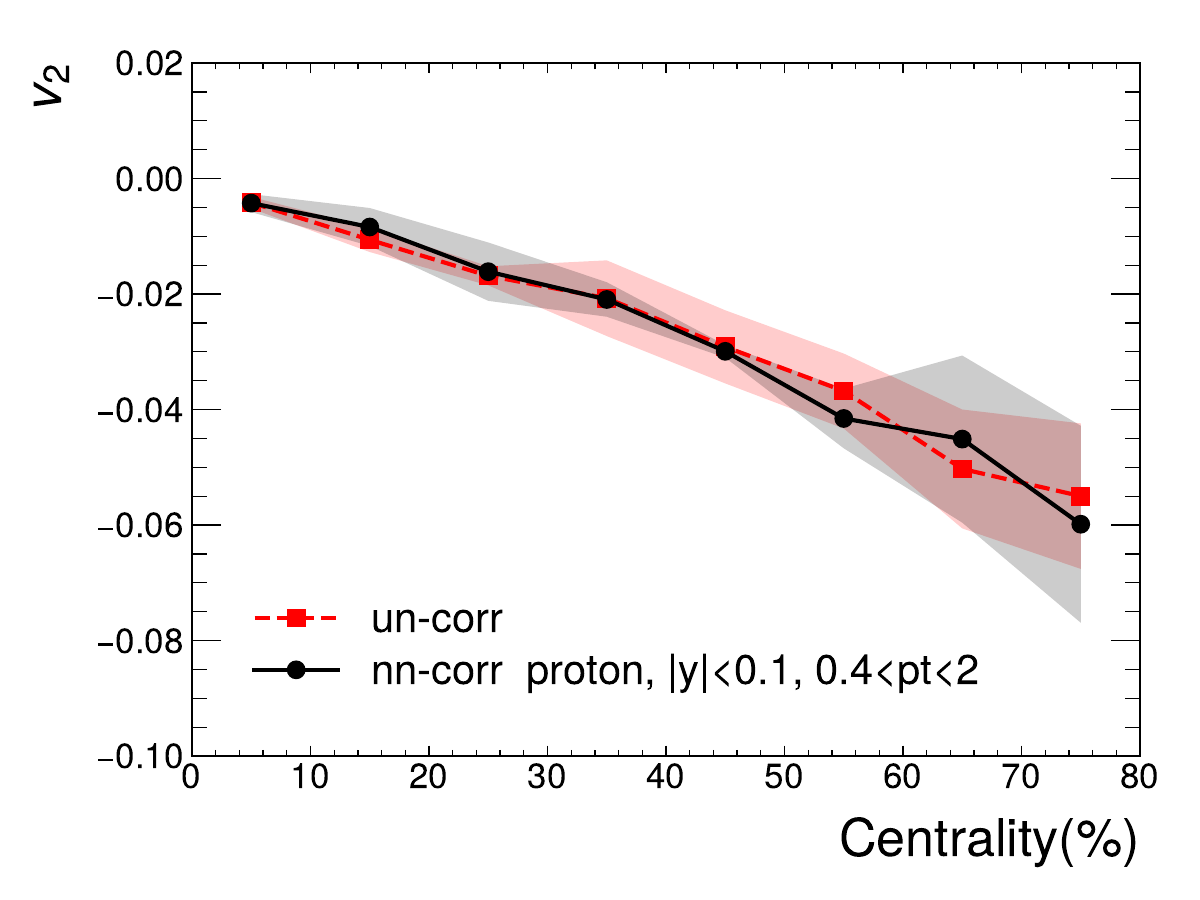}
\caption {\label{fig:v2_cent}(Color online)  The centrality dependence of $v_2$ for proton for un-corr type and nn-corr type.
}
\end{figure}

\begin{figure}[htp]
\centering
\includegraphics[width=0.48\textwidth]{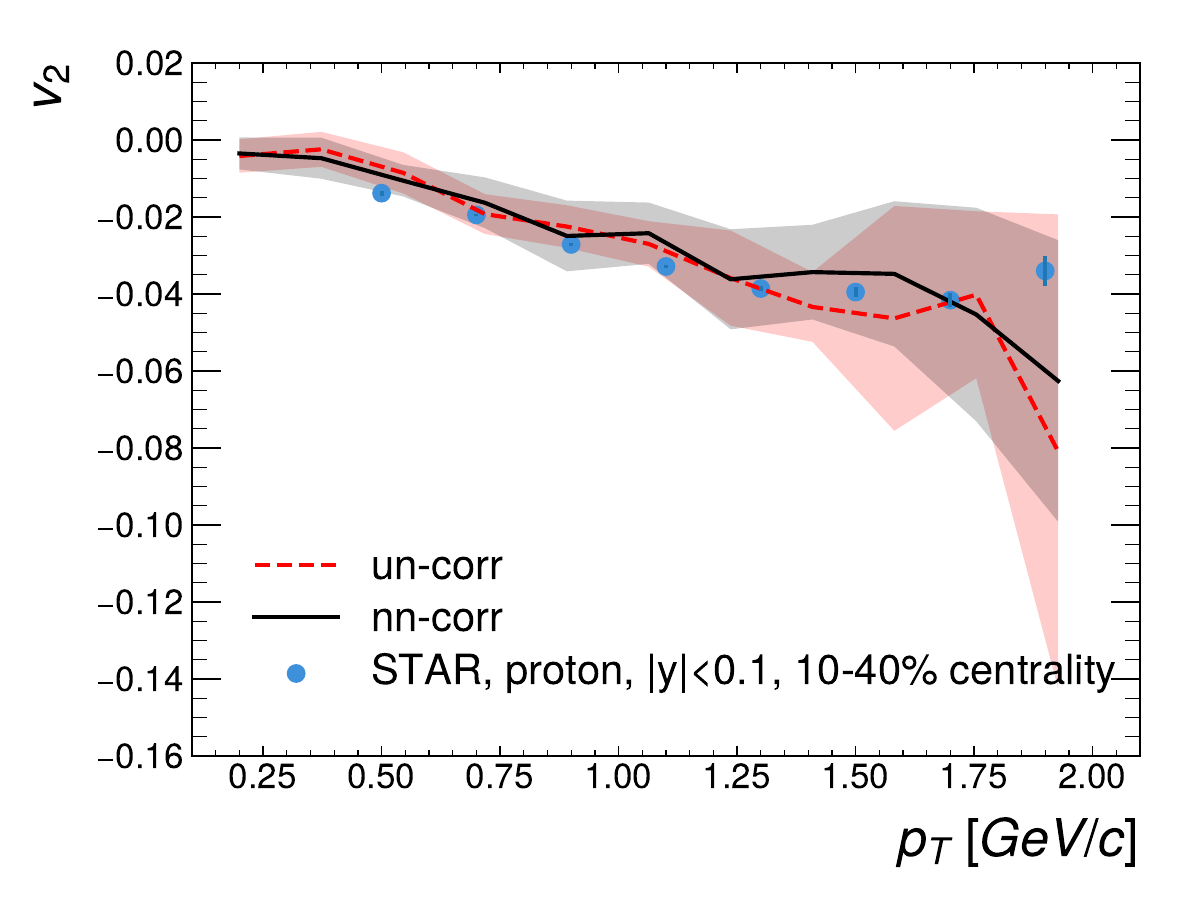}
\includegraphics[width=0.48\textwidth]{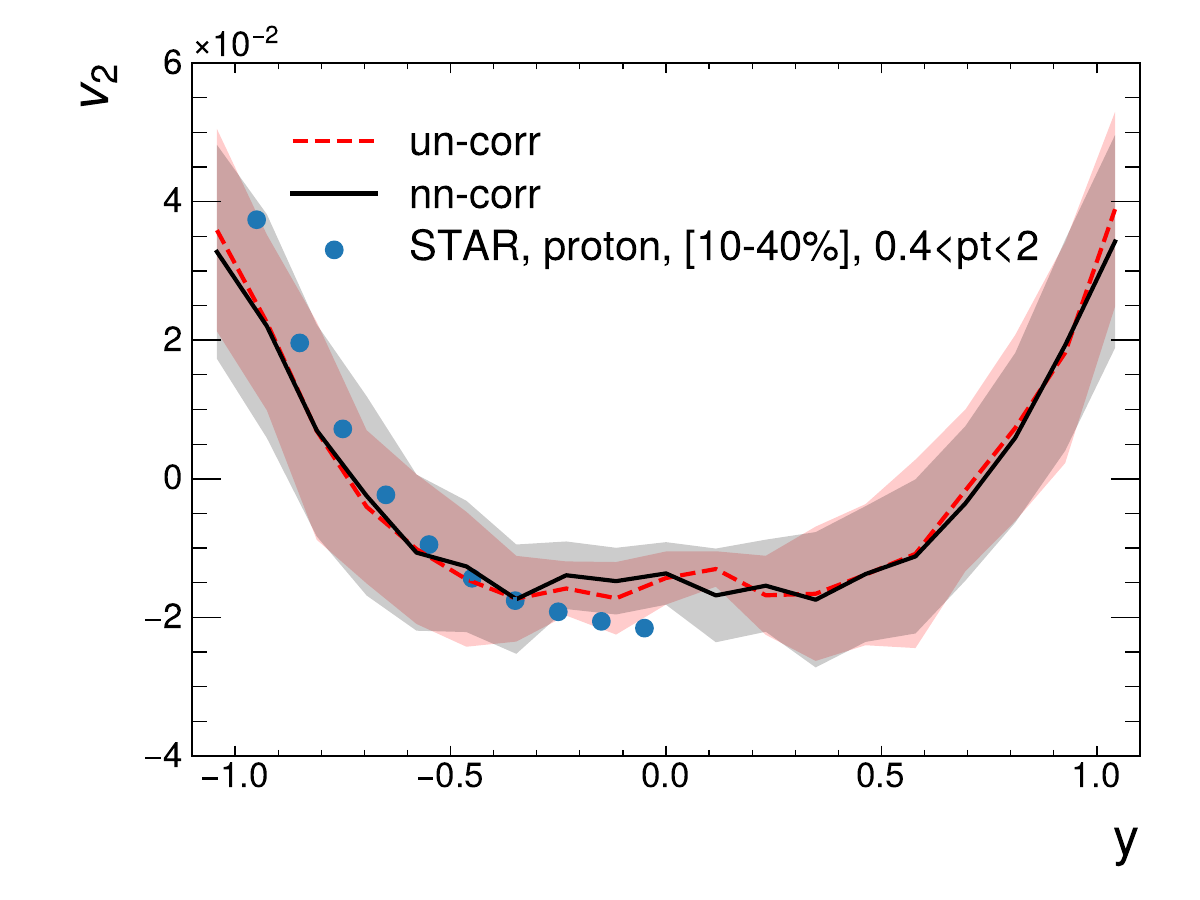}
\caption {\label{fig:v2}(Color online) The $p_T$ and y dependence of $v_2$ for protons in $^{197}\rm Au$+$^{197}Au$ collisions with 10-40\% centrality at $\sqrt{S_{\rm NN}}=3$ GeV simulated by SMASH. The experimental data from STAR Au+Au collisons at $\sqrt{S_{\rm NN}}=3$ GeV \cite{STAR:2021ozh,STAR:2021yiu}.}
\end{figure}

Fig~\ref{fig:v2_cent} and Fig~\ref{fig:v2} present the proton elliptic flow  \( v_2 \) as a function of collision centrality, transverse momentum \(p_T\) and rapidity \(y\) in the 10-40\% centrality, respectively, for direct comparison with STAR experimental data, with statistics accumulated from 150,000 simulated events.
To compute the uncertainty of model calculations, we divide the events into five equal parts and calculte $v_2$ for each part separately. 
The error bands shown in the figure are the standard deviations of these five parts. The upper figure shows \( v_2 \) plotted against \( p_T \), with protons limited to \( |y| < 0.1 \) and  0.2 GeV $< p_T <$ 2 GeV; the lower figure shows the \( y \)-dependence of \( v_2 \), with protons limited to 0.4 GeV $< p_T <$ 2 GeV. The experimental data in Fig~\ref{fig:v2} are taken from the papers \cite{STAR:2021ozh} and \cite{STAR:2021yiu}, respectively. 
When the mean field mode is activated in the SMASH model, the elliptic flow $v_2$ becomes negative, as demonstrated in the figure. The above results indicate that two-nucleon correlations in the initial nuclear structure have negligible effects on final state traditional observables.

\subsection{Identifying nucleon-nucleon correlations using Neural Network}
In our deep neural network architecture, we employ the multi-head self-attention mechanism derived from the Transformer model to process both the initial configuration and the final state generated by SMASH. Additionally, we utilize convolutional neural networks to investigate the impact of different types on the initial state entropy density distribution produced by \trento.

\begin{figure}[htp]
\centering
\includegraphics[width=0.48\textwidth]{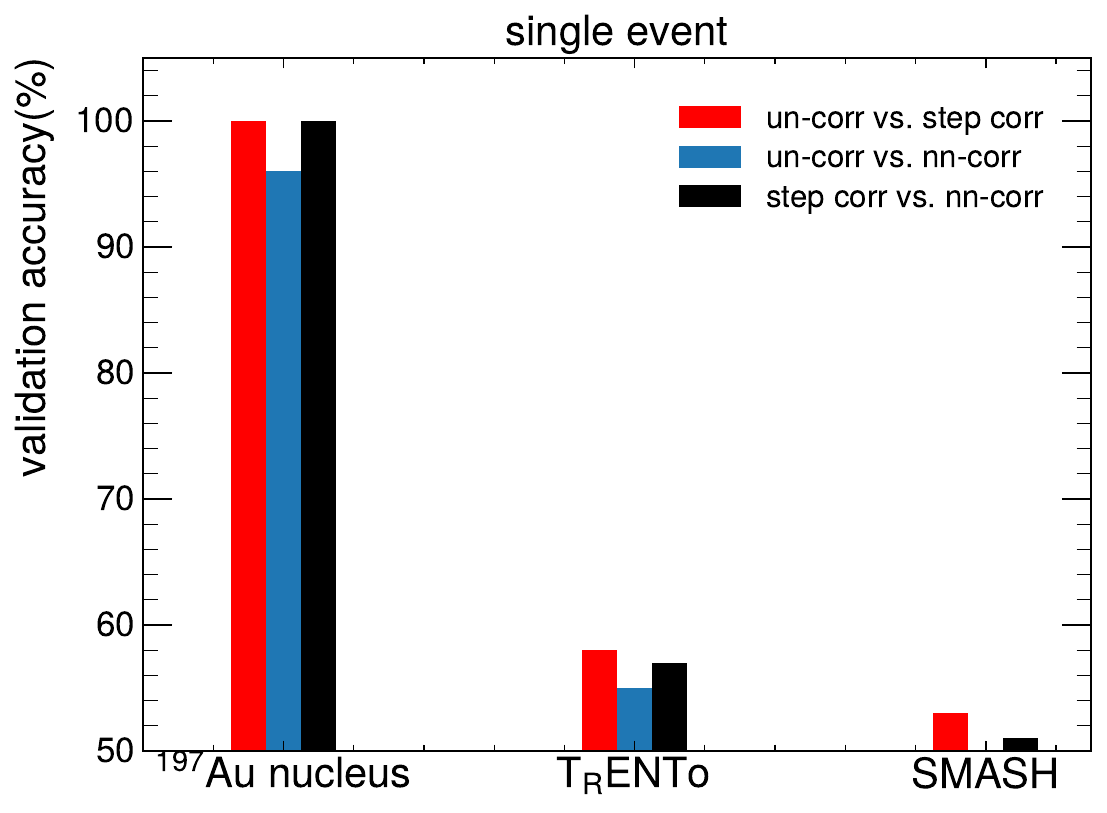}
\includegraphics[width=0.48\textwidth]{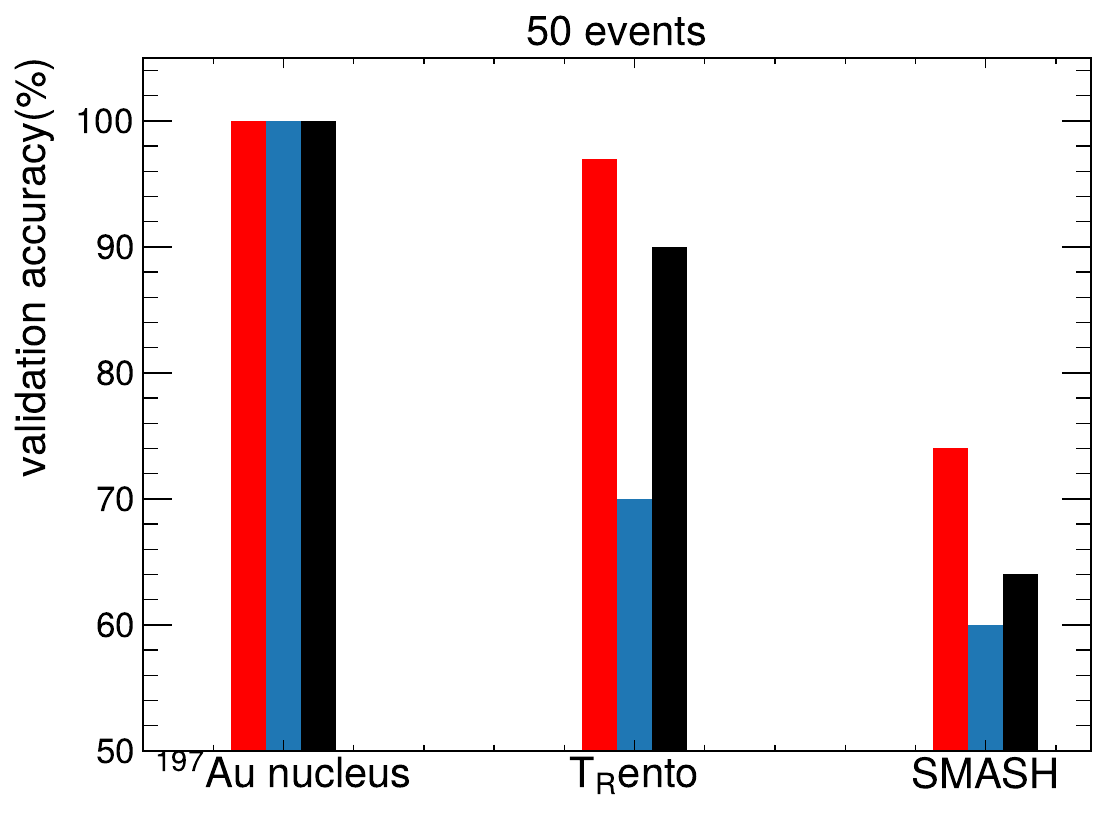}
\caption {\label{fig:nnc_acc_list}(Color online) Two-by-two classification accuracy for different types of nucleon-nucleon correlations. The top figure shows the accuracy of single event and the bottom figure shows the accuracy of 50 combined events for different types of nucleon-nucleon correlations.
}
\end{figure}

Fig~\ref{fig:nnc_acc_list} shows the two-by-two classification accuracies of the three correlation types, the deep neural network provides us with the single event and multiple events combination classification accuracies of the initial nuclei, \trento\ entropy density distribution, and final state particle cloud in momentum space from SMASH.

The left column in two sub-figures of Fig~\ref{fig:nnc_acc_list} illustrate the two-by-two classification accuracy of point cloud neural network using the initially sampled Au nucleus as input. For each Au nucleus, the spatial coordinates ($x, y, z$) of the 197 nucleons are provided. The neural network effectively distinguishes the step corr type from the other two types with accuracy approximating 100\%. However, the validation accuracy for the un-corr and nn-corr types was approximately 96\%, slightly lower than other comparisons. This discrepancy can be explained by the upper panel of Fig.~\ref{fig:nnc_initial_dist}, which shows that the difference between un-corr and nn-corr for $\Delta r < 1$ fm is smaller than their differences with the step corr. Nevertheless, after extracting latent features of each nucleon cloud and performing event mixing, the network achieves a classification accuracy of 100\%. This demonstrates that our deep learning assisted statistical analysis is capable of distinguishing subtle differences in the data. For single events, we have prepared 10,000 nucleus for each type, and found that increasing this number to 50,000 per type did not improve final classification accuracy. For event mixing, we utilized 10,000 combinations, each consisting of 50 events.

The middle column in two sub-figures of Fig~\ref{fig:nnc_acc_list} depict the pairwise classification accuracy of the deep neural network utilizing entropy density distributions generated by \trento, which typically serve as the initial conditions for relativistic hydrodynamic simulations of heavy-ion collisions. For this task, we employ deep convolutional neural networks (CNNs), as CNNs excel in image classification tasks.
Our analysis reveals that the pairwise classification accuracies for all three combinations fall below 60\% when using the initial state entropy density distribution from a single event. These accuracies are significantly lower than those obtained using the spatial coordinates of nucleons within the nucleus. This discrepancy arises because substantial information about the nuclear structure is already lost during the early stages of heavy-ion collisions, a necessary condition for achieving local equilibrium required by subsequent hydrodynamic simulations. However, by integrating a mixture of 50 events and training the CNNs on this combined dataset, the classification accuracies for these three types improve markedly to 97\%, 70\%, and 90\%, respectively. 
The event mixing is accomplished by stacking 50 entropy density distributions as 50 color channels of each image. Consistent with our earlier findings, the classification accuracy for un-corr and nn-corr remains notably lower than that of the other two groups.

The right column in two sub-figures of Fig~\ref{fig:nnc_acc_list} illustrate the performance of our novel deep neural network in binary classification on particle cloud in momentum space.
These particle clouds were 
generated from SMASH transport simulations of $^{197}\rm Au$+$^{197}\rm Au$ collisions at $\sqrt{s_{\rm NN}}=3$ GeV. Notably, the mean field mode in SMASH simulations requires significantly more computational time compared to the cascade mode. To ensure sufficient event statistics for training purposes, we employed the SMASH+cascade mode in our current simulations.
For single-event classification, the accuracy remains approximately 50\%, indicating random guessing. However, when combining 50 events, the classification accuracies improve to 74\%, 60\%, and 64\% for different scenarios. It is important to note that these classification accuracies are substantially lower than those achieved for initial state nucleus or entropy density distributions. This performance gap can be attributed to the substantial loss of nuclear structure information during the dynamical evolution of heavy ion collisions simulated by SMASH.

In contrast to our current approach, which extracts latent features from the particle cloud of each event prior to event mixing, we have also investigated the traditional multi-event mixing method. The conventional method involves computing the particle distribution in a two-dimensional space, typically the transverse momentum and rapidity space, and performs event mixing by averaging multiple images. The classification accuracy for the averaged image of 50 events using CNNs consistently remains around 51\%, significantly lower than that of our current method.

\begin{figure}[htp]
\centering
\includegraphics[width=0.48\textwidth]{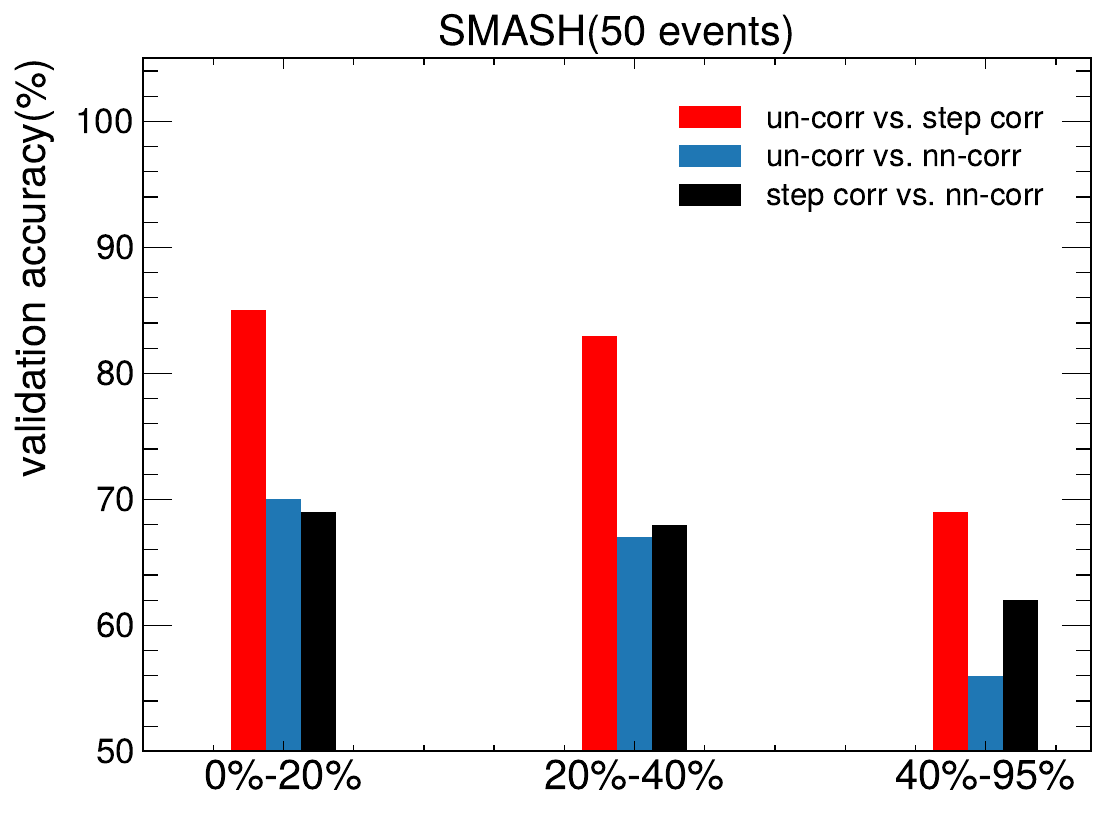}
\caption {\label{fig:nnc_acc_list_1}(Color online) Two-by-two classification accuracy for 50 combined events at different centralities in SMASH.}
\end{figure}

Fig~\ref{fig:nnc_acc_list_1} investigates the dependence of neural network classification accuracy on collision centrality. We trained a neural network using final state hadrons with centrality ranges 0\% - 20\%, 20\% - 40\%, and 40\% - 95\%, respectively. The neural network was employed to classify the nucleon-nucleon correlations into three types: un-corr, step corr, and nn-corr.
Our analysis reveals that the classification accuracy of the neural network is influenced by the centrality range. Specifically, the classification accuracies for final-state data within the 0\% - 20\% centrality range are close to that for semi-central collisions, but are significantly higher than peripheral collisions at centrality range 40\%-95\%.
We anticipated that nuclear structure information might be better preserved in peripheral collisions, as central collisions are associated with a higher degree of thermalization and a longer dynamical evolution time. Contrary to our expectations, the network demonstrates a higher classification accuracy for central collisions compared to peripheral collisions. This phenomenon can be attributed to the possibility that more information about the initial nuclear structure is encoded in central collisions. Additionally, the number of hadrons produced in central collisions is significantly larger than in peripheral collisions. However, it is important to note that the signal-to-noise ratio for central and peripheral collisions remains unknown. Our analysis indicates that our novel deep neural network is highly effective in extracting useful information from complex data.

\subsection{Interpretability of Neural Network}

\begin{figure*}[!htp]
\centering
\includegraphics[width=1\textwidth]{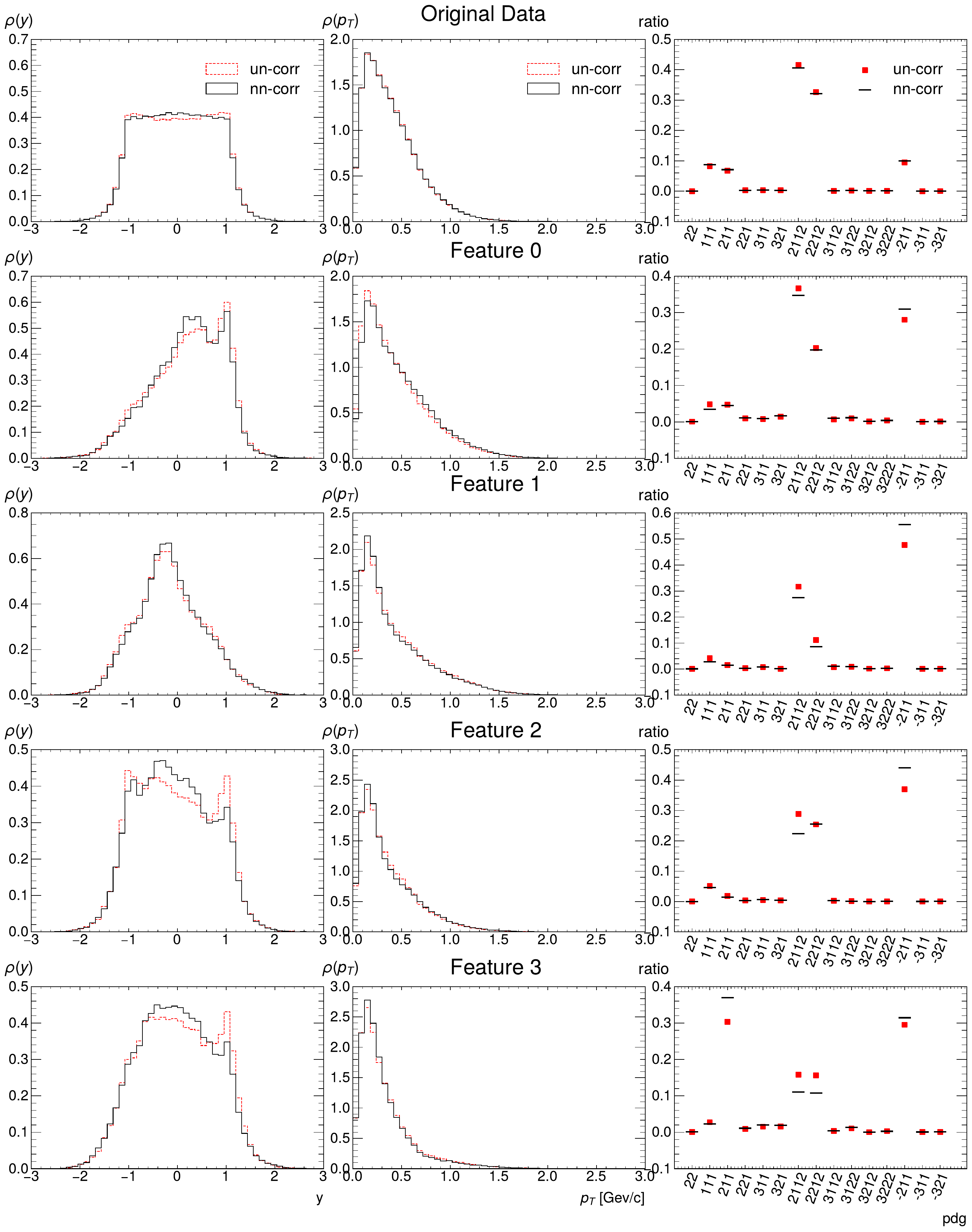}
\caption {\label{fig:nnc_feature}(Color online) Particle distribution of final state events for the un-corr and nn-corr types in different features of neutral network.
}
\end{figure*}

The interpretability of machine learning \cite{2016arXiv160204938T,2015arXiv151204150Z,2017arXiv170304730K,2020arXiv201201805M,2020arXiv200307631S} refers to the ability to discern the patterns that a deep neural network has learned to arrive at its final decision. In this section, we aim to elucidate the specific features utilized by the model to distinguish between different types of nucleon-nucleon correlations. To achieve this, we employ prediction difference analysis as our methodological approach. 

We first select correctly predicted samples with a classification accuracy above 90\% and each type has 100 groups of samples. This choice is motivated by the fact that samples with higher classification accuracy are more sensitive to the features extracted by the neural network. On the other hand, differences are not significant for features of samples with low classification accuracy. 

When a sample containing 50 events is input into the neural network, the network first transforms these 50 events into 50 sets of latent features through global average pooling. Subsequently, these features are further integrated into a single set of comprehensive features via convolution and pooling operations, as illustrated in Fig~\ref{fig:network}. Based on this processing pipeline, we are able to extract corresponding weight information at two distinct levels—individual particles and individual events—prior to the pooling operations.
Finally, we extract the top 100 particles and 10 events that are most important for classification in each event and each sample, characterized by feature values that surpass those of other particles in the cloud. We then analyze the differences in rapidity, transverse momentum, and particle composition distribution between un-corr and nn-corr for these selected samples.

Fig~\ref{fig:nnc_feature} shows the distributions of multiplicity along the rapidity direction $\rho(y)$, the transverse momentum spectra $\rho(p_T)$ and particle ratios for different nucleon-nucleon correlations, using all the selected particles that are important for decision making. The first row shows original data without feature selection. 
Based on these three distributions, it is difficult to visually distinguish between different types of nucleon-nucleon correlations. The rapidity distribution $\rho(y)$ exhibits a plateau in the central rapidity region, while both the transverse momentum spectra $\rho(p_T)$ and particle ratios show negligible variations across different correlation types.
The second row shows the effects of feature 0. All three distribution change under this feature selection. For $\rho(y)$, this feature selects particles centered at forward rapidity regions. 
For $\rho(p_T)$, the distribution is similar to the original data.
With this selection, the particle ratios between $\pi^{-}$, $n$ change significantly. 
Compared with the original data, this feature prefers events with higher multiplicity for pions than that for proton and neutron. Similar observations have been found for events selected by other features. 
The high weight particles obtained through different feature extraction exhibit significantly distinct distribution characteristics. The third row illustrates three typical distribution patterns of high weight particles extracted based on Feature 1. In the $\rho(y)$ distribution, this feature selection primarily concentrates on particles within the central rapidity region. 
For $\rho(p_T)$, the distribution becomes narrower than the original data, which selects particles trend to small transverse momentum. 
In terms of particle composition ratio, compared to the un-corr and nn-corr cases in the original data, this feature extraction significantly alters the relative proportions of $p$, $n$, and $\pi^{-}/\pi^+$, with the proportion of pions showing particularly notable enhancement. 
Not only is the difference between the un-corr and nn-corr types only visible at the particle ratio, but there are some features not shown in the figure that are significantly different from the un-corr and nn-corr types in rapidity or transverse momentum space.

The interpretability analysis of machine learning models reveals that neural networks can effectively utilize particle four-momentum information to extract meaningful physical quantities for distinguishing between the un-corr and nn-corr event types. 
Through this analysis, we observe that the network appears to define multiple novel observables by leveraging particle ratios across different momentum space regions. However, these network-derived observables prove challenging to reconstruct manually, as the final classification decision relies on complex combinations of numerous such observables.

\sect{Summary and discussion}

Our study demonstrates a novel approach for extracting information about two-nucleon distribution functions in the initial state of atomic nuclei from the final-state observables of relativistic heavy-ion collisions using deep neural networks. By developing a Monte Carlo sampling method that respects both single- and two-nucleon distributions, we generated initial states for $^{197}$Au nuclei with three distinct correlation types: un-corr, step corr, and nucleon-nucleon correlation (nn-corr) distributions derived from ab-initio calculations. The two-body density distribution for the sampled $^{197}$Au nuclei with nn-corr type showed consistency with deuteron behavior at short relative distances.

Using \trento \ initial-state model and the relativistic molecular dynamics model SMASH, we simulated Au+Au collisions at $\sqrt{S_{\rm NN}}=2.76$ TeV and $\sqrt{S_{\rm NN}}=3$ GeV. While the eccentricity distribution $\epsilon_n\{2\}$ revealed a visible reduction for nn-corr types in ultra-central \trento \  events, other observables (transverse momentum spectra, rapidity distributions, and elliptic flow) showed no significant differences between correlation types in SMASH simulations. This suggests that traditional bulk observables may be insufficient to probe initial-state nucleon correlations, motivating our machine-learning approach.

The key advancement of this work is the development of a deep neural network architecture combining point-cloud processing, self-attention mechanisms, and latent-space correlation analysis. This framework successfully distinguished nn-corr from un-corr configurations using final-state momentum-space data, with several notable findings: (1) Classification accuracy was highest when using initial-state nucleons and decreased progressively for \trento \  and SMASH outputs, reflecting information loss during pre-equilibrium and dynamical evolution. (2) The network more easily discriminated step-corr from un-corr than the other two combinations, suggesting nn-corr introduces subtler signatures. (3) Central collisions yielded better discrimination than semi-central events, likely due to more correlated nucleon pairs. (4) Our method significantly outperformed traditional multi-event mixing approaches (accuracy $\sim$ 51\%) that average particle distributions.

Interpretability analysis revealed that the network constructs physically meaningful observables through nonlinear combinations of features, with high-weight particles exhibiting distinctive momentum-space patterns and hadron ratios. This contrasts with conventional analyses that typically examine one-dimensional projections of phase-space distributions.

Our current model considers only spatial correlations, neglecting momentum correlations that account for ~70\% of nucleon kinetic energy in the nuclear rest frame \cite{Hen:2016kwk}. Developing a momentum-aware sampling method could be particularly impactful for low-energy collision studies.
As highlighted by \cite{Huang:2023viw}, n-body correlations beyond two-nucleon distributions play significant roles in light nuclei like $^{16}$O. While we focused on two-nucleon effects in heavy nuclei, extending this framework to three-nucleon correlations presents a natural next step. The decreasing classification accuracy through successive stages of dynamical evolution, from initial state to \trento\  and SMASH, suggests that early-time observables may offer better sensitivity to initial-state correlations than final-state measurements.

\begin{acknowledgments}
This work is supported by the National Natural Science Foundation of China under Grant No.\ 12075098, No.\ 12435009 and No.\ 1193507, and the Guang-dong MPBAR with No.2020B0301030008. The numerical calculation have been performed on the GPU cluster in the Nuclear Science Computing Center at Central China Normal University (NSC3). 
\end{acknowledgments}

\bibliography{ref}

\end{document}